\documentclass[journal]{IEEEtran}
\usepackage{amsmath,amsfonts}
\usepackage{algorithmic}
\usepackage{array}
\usepackage[caption=false,font=normalsize,labelfont=sf,textfont=sf]{subfig}
\usepackage{textcomp}
\usepackage{stfloats}
\usepackage{url}
\usepackage{verbatim}
\usepackage{graphicx}
\usepackage{cite}

\usepackage{multirow}
\usepackage[ruled,vlined]{algorithm2e}

\usepackage{xcolor}

\begin{document}

\title{RecFlash: Fast Recommendation System on In-Storage Computing with Frequency-Based Data Mapping}

\author{
    \IEEEauthorblockN{   
        Jangho Baik\IEEEauthorrefmark{2}\textsuperscript{*},
        Sunghyun Kim\IEEEauthorrefmark{2}\textsuperscript{*},
        Gisan Ji\IEEEauthorrefmark{2},
        Wonbo Shim\IEEEauthorrefmark{3},
        and~Sungju Ryu\IEEEauthorrefmark{4},~\IEEEmembership{Member,~IEEE}
        }       
        
    \IEEEauthorblockA{\IEEEauthorrefmark{2}Sogang University, Seoul, Republic of Korea}  
    
    \IEEEauthorblockA{\IEEEauthorrefmark{3}Seoul National University of Science and Technology, Seoul, Republic of Korea}
    
    \IEEEauthorblockA{\IEEEauthorrefmark{2}\{janghobaek, sunghyun, gisanji, sungju\}@sogang.ac.kr , \IEEEauthorrefmark{3}wbshim@seoultech.ac.kr}

\thanks{This work is an extended version of a paper presented at the IEEE APCCAS 2025 \cite{baik2025recflash}. Compared to the conference version, this manuscript includes online training support with adaptive remapping, extended hardware implementation details, and a broader evaluation on real datasets. Details are summarized in Appendix \ref{appendices}.\vspace{0.3em}}
\thanks{\textsuperscript{*}Equal contribution.\vspace{0.1em}}
\thanks{\IEEEauthorrefmark{4}Corresponding Author}
}






\maketitle
\begin{abstract}
Recommendation system has gained a large popularity for a variety of personalized suggestion tasks, but the ever-increasing number of user data makes real-time processing of recommendation systems difficult. 
NAND flash memory-based in-storage computing scheme can be one of favorable candidates among the various acceleration approaches because the flash memory typically has a larger memory capacity than the other memory types, so it can efficiently handle a large amount of user data for the recommendation inference services.
However, different from other neural network applications where data is sequentially fetched from memory, the recommendation system shows the irregular random memory access pattern.
Hence, most of the data loaded from the NAND flash array to the page buffer are not used, so a large portion of the internal bandwidth is underutilized, which degrades the performance on the inference acceleration of the recommendation tasks.
In this paper, we propose RecFlash, a fast recommendation inference accelerator utilizing a data remapping algorithm with NAND flash-based in-storage computing (ISC).
The experimental results show that our proposed method improves the latency and energy consumption by up to 81\% and 91.9\%, respectively, over the existing NAND flash-based ISC architecture.
\end{abstract}

\begin{IEEEkeywords}
Recommendation system, NAND flash memory, in-storage computing, hardware accelerator, data remapping.
\end{IEEEkeywords}
\section{Introduction}
\label{section_intro}

\IEEEPARstart{R}{ecommendation} systems provide personalized suggestions by analyzing data such as streaming history, clicks, and social network interactions. 
Companies like Amazon, eBay, and Alibaba boost sales with relevant product recommendations, while platforms like YouTube, Netflix, and Meta enhance competitiveness with tailored videos and news feeds. 
Collaborative filtering \cite{he2017neural} assumes that past experiences determine future preferences, while content-based filtering \cite{mooney2000content} analyzes correlations between items and user profiles. 
DNN-based recommendation systems can handle complex non-linear relationships between users and items, providing feature combinations beyond traditional methods. The DLRM \cite{naumov2019deep} uses sparse-length sum (SLS) operations to convert categorical features into continuous ones. Embedding layers retrieve vectors from lookup tables (LUTs), perform pooling to encode user history, and process data through fully connected layers. 

However, DNN-based systems require significant computational resources. 
In typical data centers, over 80\% of hardware is dedicated to personalized recommendation services. 
In Meta’s data centers, DLRM inferences are performed trillions of times daily \cite{hazelwood2018applied,kim2023recpim}.
While the embedding layer itself has low computational intensity, it demands a substantial memory footprint and shows random access patterns to the memory.
To relieve the Von Neumann bottleneck between processor and memory, many DRAM-based near-memory computing approaches have been explored.
TensorDIMM \cite{kwon2019tensordimm} introduced a near-memory processing method which increased the effective bandwidth for the embedding layers by placing the SLS units on the buffer chip in the DIMM.
RecNMP \cite{ke2020recnmp} performed in-depth workload characterization which improves the cache hit ratio by exploiting temporal reuse patterns. 
By doing so, the RecNMP reduced the number of memory accesses for embedding lookup operations, and it obtained significant SLS speedup in the DLRM workload.
SPACE \cite{kal2021space} proposed the heterogeneous near-memory processing architecture. 
It utilizes the two types of hierarchies to balance the bandwidth between the HBM and DRAM, and it also exploited the locality of the partial sum for the embedding layers.
RECROSS \cite{liu2023accelerating} improves the effective bandwidth by parallelizing the computation on multiple memory hierarchies including rank/bank-group/subarray-level parallelism. 
Using the proposed partitioning technique for the embedding table, the tiles are assigned to corresponding NMP-levels for the efficient parallel computing.

\begin{figure}[!t]
\centering
\includegraphics[width=0.48\textwidth]{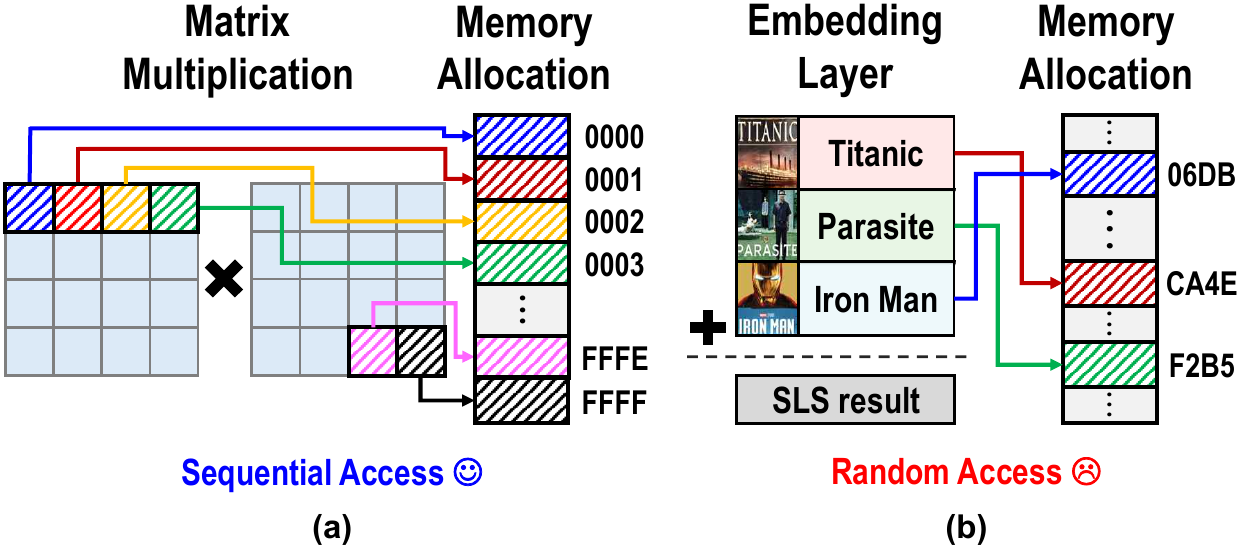}
\vspace{-2mm}
\caption{Memory access patterns on (a) general matrix multiplication and (b) embedding layer for recommendation system.}
\vspace{-6mm}
\label{intro_access}
\end{figure}

Meanwhile, the data size for the personalized recommendation system has been exponentially growing.
As embedding tables grow in size, the recommendation inference services often consume more than hundreds of gigabytes of the storage.
For example, the recent Criteo Click Logs dataset is more than tera-byte level \cite{criteo-terabyte}, and it cannot be handled in the DRAM-level \cite{wilkening2021recssd,sun2022rm}.
As a result, the processor must frequently access the lower level disk to obtain a large number of parameters, which leads to a large latency for the access to the disk.
The NAND flash memory offers faster read/write speed and higher bandwidth compared to the traditional hard-disk-based storage solutions.
To exploit the advantage of storage density with the NAND flash memory, several approaches to NAND flash-based in-storage computing (ISC) have been introduced.
RecSSD \cite{wilkening2021recssd} enhances the performance of SSD-based recommendation systems by processing embedding lookups in the SSD, while RM-SSD \cite{sun2022rm} uses a low-cost FPGA to offload the recommendation system to the SSD and optimize MLP layer processing.
Meanwhile, during the read operation of the NAND flash memory, data stored in the array is first fetched to the page buffer. Afterward, the buffered data is sent to the outside memory chip.
In recommendation inference tasks, embedding vectors range from 64 to 512B, while the page buffer size in commercial single-level cell (SLC) NAND flash is 4KB \cite{sun2022rm}, much larger than the embedding vector size.
In the recent multi-level cell arrays (eg., MLC, TLC, QLC) \cite{khakifirooz202130,park202130,kim202328,huh202013}, the page buffer size increases (8KB-32KB).
Note that different from other neural network workloads where data is sequentially loaded from memory, SLS operation in the embedding layer requires the random memory access pattern.
Only a part of the data loaded to the page buffer from memory are used, and most of the data cannot be utilized, because embedding vectors are much smaller than the page buffer.
As a result, most of the internal memory bandwidth in the NAND flash is underutilized, which degrades the system performance.

To relieve such a limitation of the recommendation system inference in the NAND flash-based ISC, we maximize the effective internal memory bandwidth for the SLS computations by introducing a data remapping scheme and the page-wise cache for the embedding tables.
The key observations of our work \cite{baik2025recflash} are summarized as follows.

\begin{enumerate}\itemsep5pt

\item We address and analyze the bandwidth under-utilization issue in the NAND flash-based ISC on the personalized recommendation system inference tasks.

\item We present the access frequency-based data remapping scheme of the embedding vectors to improve the reusability of the data stored in the page buffer of the NAND flash memory.

\item We introduce a page-wise cache which maximizes the reusability of the embedding vectors with a page-wise least recently used (LRU) replacement policy. 

\item Based on the observation above, we introduce the NAND flash-based ISC architecture called RecFlash which achieves fast and energy-efficient personalized recommendation inference acceleration.

\end{enumerate}


\section{Preliminaries}
\label{section_prelim}

\subsection{Random Data Access on Recommendation System}
\label{section_prelim_random}
Most of the conventional neural network layers such as matrix-matrix multiplication, matrix-vector multiplication, convolution, and attention typically show the sequential memory access pattern.
For example, in the matrix-vector multiplication, the elements of the input matrix and vector are stored in the consecutive memory addresses (Fig. \ref{intro_access}a).
On the other hand, in the recommendation system, the user-item interaction data between users and items is typically very sparse.
In other words, a user may not provide ratings for all the items, and only a part of the items is connected with the user.
Due to the sparsity, in the embedding layers, embedding vectors are stored in irregular row numbers of the embedding lookup table.
Fig. \ref{intro_access}b illustrates an example of the random access pattern in the embedding lookup stages.
If the user has watched \textit{Titanic}, \textit{Parasite}, and \textit{Iron Man} among the a bunch of items in the movie category, only required embedding vectors are located at rows CA4E/F2B5/06DB.
After we fetch these embedding vectors for multiple categories, the SLS computation is performed thereby generating a sum value for each category.

\begin{figure}[!t]
\centering
\includegraphics[width=0.48\textwidth]{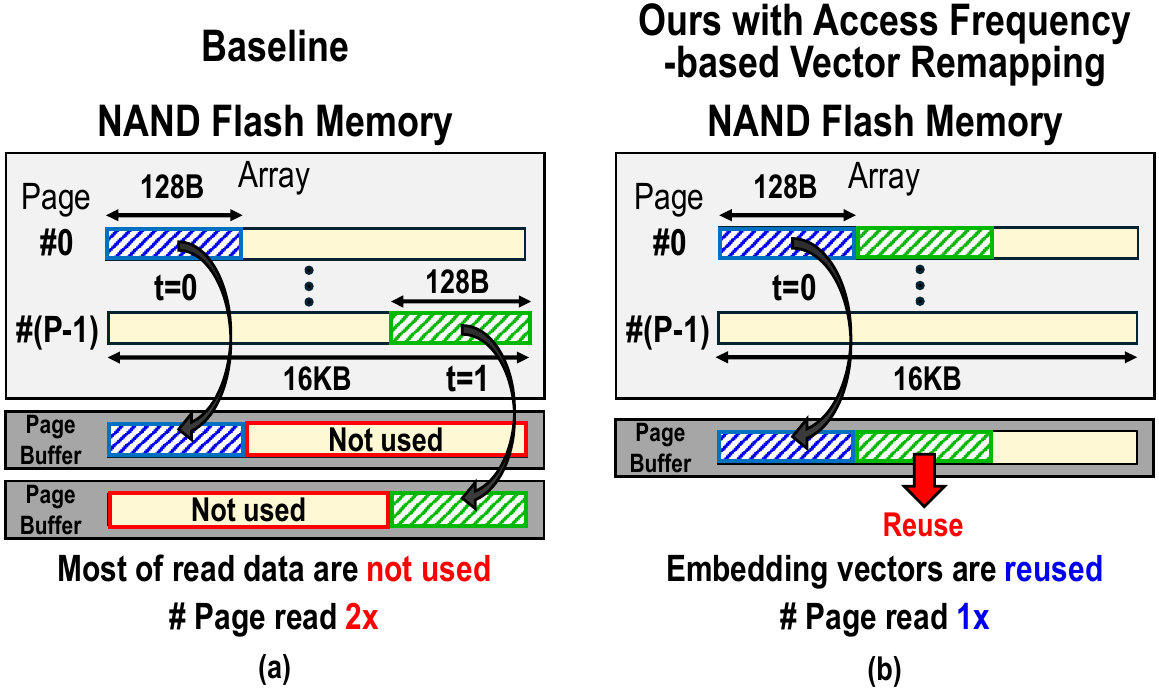}
\vspace{-2mm}
\caption{(a) Inefficient bandwidth utilization on NAND flash with baseline mapping method and (b) our method with remapping embedding vectors. We assume that an embedding vector size is 128B, and a page buffer size is 16KB.}
\label{prelim_bandwidth}
\end{figure}
\begin{figure}[!t]
\centering
\includegraphics[width=0.48\textwidth]{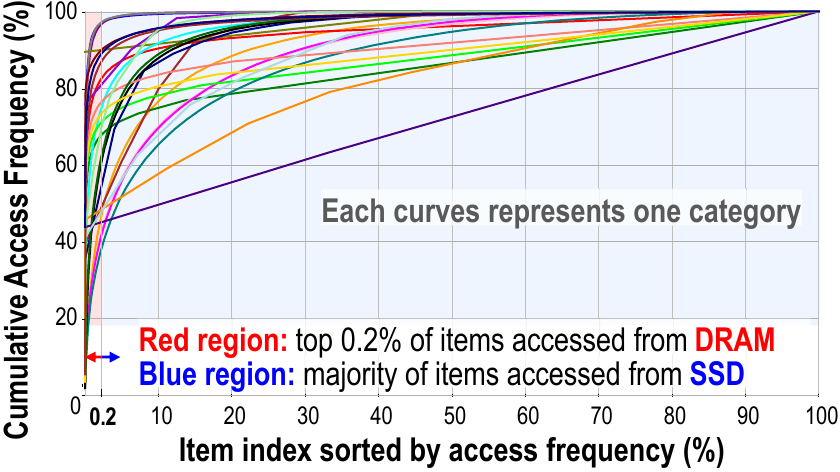}
\vspace{-2mm}
\caption{Data access frequency in the embedding layer on Criteo TB dataset.}
\vspace{-4mm}
\label{prelim_freqency}
\end{figure}
\subsection{Inefficient Bandwidth Utilization on NAND Flash Memory}
\label{section_prelim_page}
For the memory read operation, a page row is selected, and the data in the row is fetched to the page buffer located at the bottom of the array.
In the typical matrix multiplication, input/weight data are stored in the consecutive indices of the memory.
So, once a page is loaded to the page buffer, all the values in the page buffer are sequentially used for the matrix multiplication.
On the other hand, please remember that embedding layers show the random access pattern for the embedding lookup as explained in the Section \ref{section_prelim_random}.
As a result, only a few data in the page buffer is used for the SLS computation and the other values are mostly not used (Fig. \ref{prelim_bandwidth}a).
Considering that an embedding vector size is 64-512B and the page buffer size of is 2-32KB, a large amount of the internal bandwidth between flash array and the page buffer is abandoned, so such an under-utilization of the memory bandwidth leads to significant performance degradation.
Furthermore, recent flash memory with multi-level cells including MLC/TLC/QLC has a large number of latches in a page buffer than SLC array, so the loss of the throughput and energy becomes much more severe than the previous SLC case.

Meanwhile, it is widely known that the recommendation system has temporal locality \cite{ke2020recnmp,wilkening2021recssd}.
Users tend to frequently access the preferred items than others.
For instance, most gamers prefer a few popular games, while a small number of users try to experience a variety of games.
To deal with temporal locality, previous studies have used DRAM caches to store frequently accessed embedding vectors.
For example, RecSSD stores a portion of the embedding vectors from each embedding table in the DRAM cache. However, with the Criteo TB dataset, only 0.2\% of the total embedding vectors are stored in the cache, leading to cache misses when accessing the remaining vectors (Fig. \ref{prelim_freqency}). In the case of the cache miss, we try to access the flash memory array with the degraded intra-flash bandwidth, which largely reduces the inference latency.

\begin{figure*}[!t]
\centering
\includegraphics[width=1.0\textwidth]{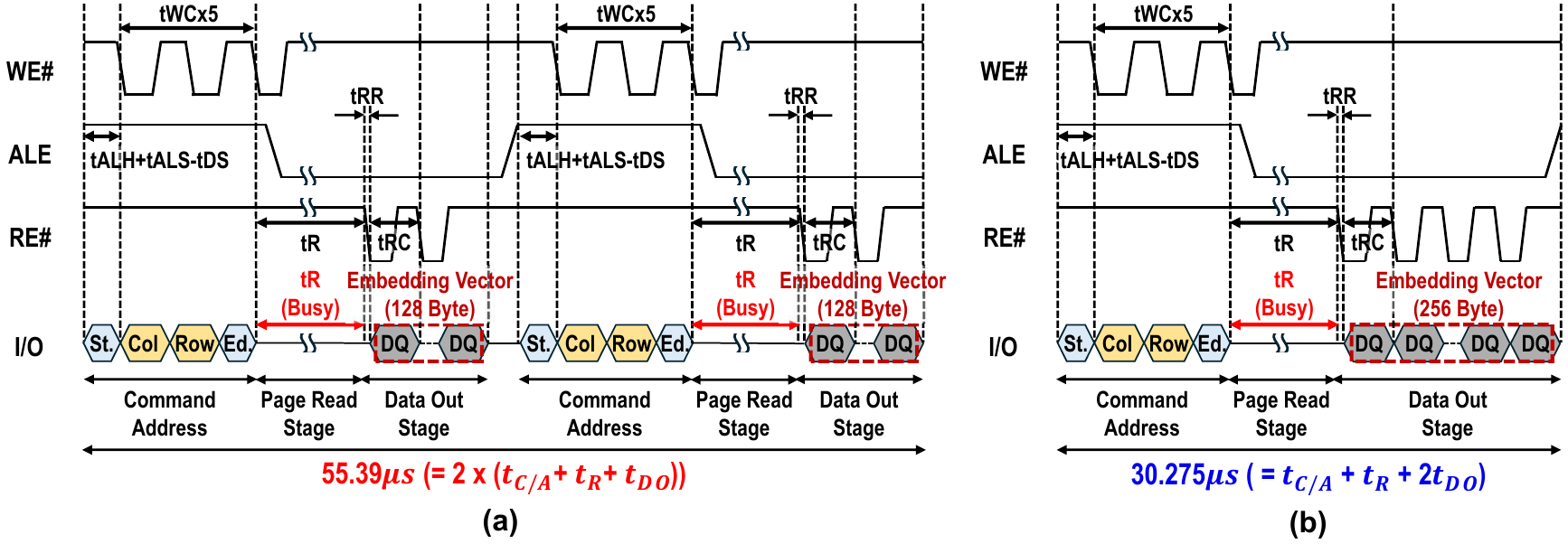}
\vspace{-6mm}
\caption{Timing diagram of read operation for 2 embedding vectors in the NAND flash memory. (a) 2 embedding vectors are located at 2 different pages. (b) 2 embedding vectors are located at 1 page. WE\#: Write Enable. ALE: Address Latch Enable. RE\#: Read Enable. I/O\# : Data I/O port. Parameters are listed in Table \ref{table_sparsity}.}
\vspace{-6mm}
\label{proposed_timing}
\end{figure*}

\begin{table}[!t]
\renewcommand{\arraystretch}{1.2}
\caption{The Timing Parameters Used for the Analysis on the Read Operation of NAND Flash \cite{ltc3600}.}
\centering
\footnotesize 
\begin{tabular}{c c c}
\hline
\textbf{Symbol} & \textbf{Description} & \textbf{Time [\boldmath{$\mu$s}]} \\
\hline
$t_{\text{ALH}}$ & ALE hold time & 0.005 \\ \hline
$t_{\text{ALS}}$ & ALE setup time & 0.01 \\ \hline
$t_{\text{DS}}$ & Data setup time & 0.007 \\ \hline
$t_{\text{WC}}$ & Write cycle time & 0.02 \\ \hline
$t_{\text{R}}$ & Data transferring from array to buffer & 25 \\ \hline
$t_{\text{RR}}$ & Ready to RE\# falling edge & 0.02 \\ \hline
$t_{\text{RC}}$ & Read cycle time & 0.02 \\
\hline
\end{tabular}
\vspace{-6mm}
\label{table_sparsity}
\end{table}

\section{Proposed RecFlash Architecture}
\label{section_proposed}

In the previous Section, we observed that embedding operations show the random access pattern, so only a few bits are used in a single page read operation, and most of the data are not used, which significantly reduces the performance.
To mitigate the problem, we propose a RecFlash hardware architecture in this Section.
The key idea of our work is to analyze the access pattern with the sampled training data and to perform the remapping of the embedding values in the flash memory array before the weight training stage.

\subsection{Motivation: Read Operation of NAND Flash Memory}
\label{section_proposed_motivation}

\begin{figure*}[!t]
\centering
\includegraphics[width=0.98\textwidth]{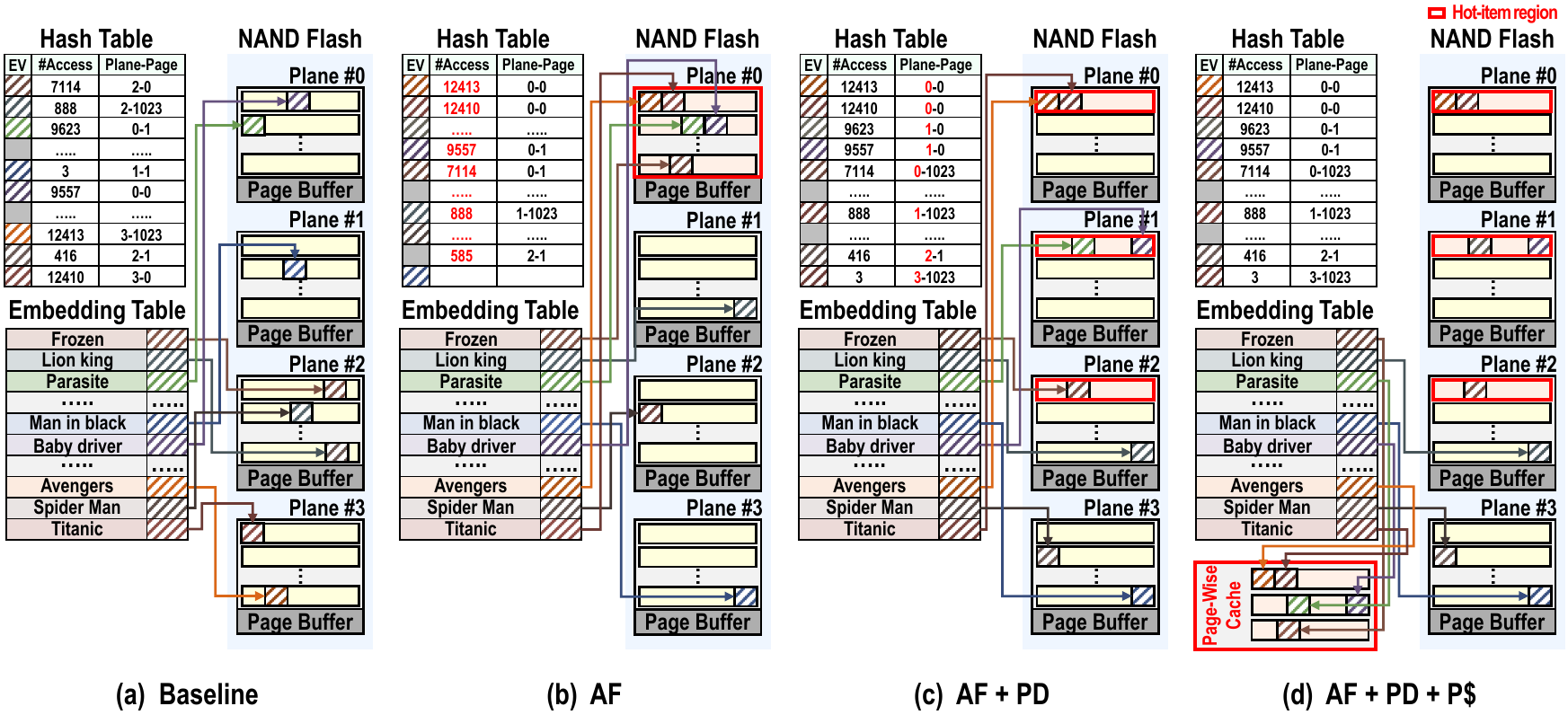}
\vspace{-2mm}
\caption{Proposed RecFlash method for the acceleration of recommendation system. (a) Baseline mapping method on NAND flash. Proposed mapping methods with (b) AF, (c) AF+PD, and (d) AF+PD+P\$ where AF: access frequency-based remapping, PD: plane distribution, and P\$: page-wise cache.}
\vspace{-6mm}
\label{proposed_mapping}
\end{figure*}
Fig. \ref{proposed_timing} shows the comparison of the read operation between the baseline and our work.
We assume that we require 2 embedding vectors for the SLS computation for the explanation.
Considering that the embedding layer shows the random access pattern, 2 embedding vectors are mostly stored in the different rows of the memory array (Fig. \ref{prelim_bandwidth}a).
Therefore, to perform the SLS operation, we need to execute 2 independent page read operations (Fig. \ref{proposed_timing}a).
The read operation of the flash memory consists of 3 stages: command address (C/A) stage, page read stage, and data out stage.
The C/A stage starts with the command for the column/row access (St.), which is followed by the column address (Col).
Then, the controller asserts the row address (Row).
If the bit-width for the address pin is limited, the column/row addresses are divided into a few parts, and each part is sequentially fed to the decoder.
The time for the C/A stage is $t_{C/A}$ (Eq. \ref{eq1}).
\begin{equation}\label{eq1}
t_{C/A} = (t_{ALH} + t_{ALS} - t_{DS}) + t_{WC} \times 5 + t_{DS}
\end{equation}
After the C/A stage is complete (Ed.), the page read stage is performed.
In the stage, we access the page in the memory array with the column/row addresses.
The data stored in the selected page are read through the strings and then they are finally latched by the page buffer.
The time for the page read stage is $t_{R}$. Next, in the data out stage, the data from the page buffer are transferred outside the NAND flash chip through the I/O pins.
The time for the data out stage is $t_{DO}$ (Eq. \ref{eq2}) where $N=\#Data$ to be fetched.
\begin{equation}\label{eq2}
t_{DO} = t_{RR} + t_{RC}  \times N
\end{equation}

Table \ref{table_sparsity} shows the detailed parameters for the read operation.
Using the parameters, $t_{C/A}=0.115\mu s$, $t_{R}=25\mu s$, and $t_{DO}=2.58\mu s$.
If we load 2 embedding vectors from multiple pages, we require $55.39\mu s (=2\times(t_{C/A}+t_{R}+t_{DO}))$.

In contrast, our approach performs the remapping of embedding vector addresses (Fig. \ref{proposed_timing}b).
Based on the access frequency, we perform the remapping embedding vectors on the memory arrays (Fig. \ref{prelim_bandwidth}b, The method will be explain in the following Section).
Then, multiple embedding vectors in a page buffer can be placed in the same page, and they are reused without loading another page row from the array.
Suppose that we need 2 embedding vectors in the same manner as baseline, the required time is reduced to $30.275\mu s(=t_{C/A} + t_{R}+2t_{DO})$.
There may be a concern that the timing overhead is required to rearrange the address of the entire embedding vector table in real-time.
However, we do not perform such a heavy {\fontfamily{qcr}\selectfont sort} operation for the entire embedding table in the real-time, and
we only sort a part of the embedding table, which consists of \textbf{hot items}.
As a result, the overhead is relatively small compared to the inference time
because we perform the hash-table-aware remapping before the training stage without training-time overhead (Section \ref{section_proposed_recflash_remapping}) and we achieve low-overhead adaptive mapping in the real-time online training (Section \ref{section_proposed_recflash_online}).

\subsection{Embedding Layers in Previous NAND Flash-based Accelerators}
\label{section_proposed_prev}

RecSSD sequentially reads the data loaded into the page buffer to retrieve the necessary data. However, if the required data is located at the end of the page buffer, the sequential read characteristic necessitates reading all the unnecessary data before it, which can increase $t_{DO}$. Meanwhile, RM-SSD minimizes access latency by selectively reading data from the page buffer of $t_{DO}$ stage.
However, as explained in Section \ref{section_proposed_motivation}, $t_{DO}$ accounts for only a small part of the read latency, and the random access pattern of the embedding layer leads to an increase in $t_{R}$, resulting in longer overall read operation times.
Therefore, RM-SSD's approach does not show dramatic improvement of overall performance in the embedding layers.
In Section \ref{section_result}, we will show the performance improvement of our design over those SSD-based previous works.
\subsection{RecFlash Approach}
\label{section_proposed_recflash}

\subsubsection{Access Frequency-based Remapping}
\label{section_proposed_recflash_remapping}

Fig. \ref{proposed_mapping}a shows the baseline approach for mapping embedding tables to memory. In flash memory, a {\fontfamily{qcr}\selectfont block} contains multiple {\fontfamily{qcr}\selectfont page} rows, a {\fontfamily{qcr}\selectfont plane} (array) has multiple blocks, and each plane has a dedicated {\fontfamily{qcr}\selectfont page buffer}. As explained in Section \ref{section_prelim_random}, embedding layers have a sparse access pattern, with \textbf{hot items} (frequently accessed items) scattered across various pages and planes. Since only a few vectors are used in the page buffer, much of the internal bandwidth between the array and page buffer is wasted.
To increase the utilization of data in the page buffer, \textbf{hot items} can be gathered into a single page (Fig. \ref{proposed_mapping}b)) instead of being scattered.
First, we analyze access counts for the entire lookup table, sort the embedding vectors in descending order based on their access frequency, and sequentially place the sorted vectors into each page of the first plane. Once the placement in the first plane is complete, we continue assigning physical addresses to the next plane in the same manner as the first plane.
However, this leads to \textbf{hot items} being clustered in only a few planes, while the other planes contain mostly cold items, which are rarely used. Since each plane has its own page buffer, only a few buffers are active, significantly reducing bandwidth utilization in the planes.
Instead, we distribute \textbf{hot items} across multiple planes (Fig. \ref{proposed_mapping}c). 
This approach enables plane-level parallelism,
allowing more page buffers to be active and increasing throughput for embedding lookups. There may be concerns about latency overhead from remapping embedding vectors, but we mitigate this by not directly sorting the embedding tables. In the baseline approach (Fig. \ref{proposed_mapping}a), a hash table links embedding vector addresses with physical flash memory addresses. In our design (Fig. \ref{proposed_mapping}b-c), we assign physical addresses to vectors in descending order based on access frequency when creating the hash table. To minimize preprocessing time, we use a sampled training set for access count analysis.
So far, such a remapping is completed before the training phase, so the additional overhead is negligible during training and does not increase inference latency.

\subsubsection{Page-Wise Cache}
\label{section_proposed_recflash_cache}

In computer architecture, a cache stores frequently accessed data in memory for reuse. In a NAND flash-based recommendation system, when multiple embedding vectors are grouped at the page level, it is more efficient to operate the cache at the page level.  
Therefore, we implemented a page-wise cache by placing a 128KB SRAM in the SSD controller which occupies an area of 0.44mm\textsuperscript{2} in the 28nm CMOS technology, accelerating embedding operations.
The page-wise cache can be stored inside NAND flash memory, but we use vanilla NAND flash memory without modifying the commercial design. Instead, we implement the page-wise cache in the SSD controller chip.
Fig. \ref{proposed_mapping}d illustrates an SSD architecture that stores page-level data, including \textbf{hot items}, in an SRAM-based cache. If an embedding vector is in the cache, it is retrieved directly without reading from the NAND flash. 
In the case of a cache miss, the page is loaded from the NAND flash into the buffer, and the data is stored in the cache. 
The cache is managed by the LRU replacement policy, reducing the frequency of page buffer accesses and improving the efficiency of embedding operations.
\begin{figure}[!t]
\centering
\includegraphics[width=0.48\textwidth]{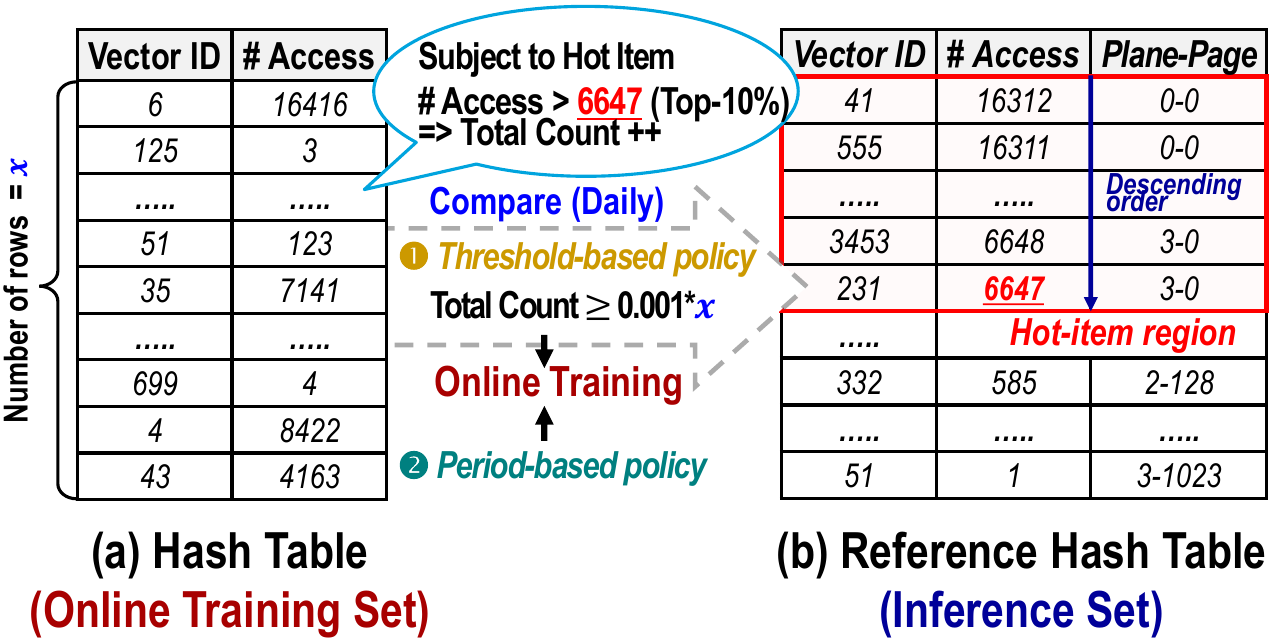}
\vspace{-2mm}
\caption{(a) Hash table recording access counts of embedding vectors observed during online inference. (b) Reference hash table constructed from sampled training data and sorted in descending order based on access frequency.}
\vspace{-4mm}
\label{Online_Training_Begin}
\end{figure}
\subsubsection{Online Training Condition}
\label{section_proposed_recflash_condition}

So far, we supposed that the embedding vectors are pre-trained before the inference service, but
industrial-scale recommendation systems require online training which updates the embedding table in real-time as well as using pre-trained data at the offline stage \cite{hazelwood2018applied,wang2024rap,lai2023adaembed,bother2023towards}.
Various trigger conditions have been proposed to determine when to perform online training of the recommendation system \cite{lai2023adaembed,bother2023towards}.
AdaEmbed \cite{lai2023adaembed} proposes a \textit{threshold-based trigger policy} that computes importance scores based on the access frequency of embedding vectors and adjusts the training schedule accordingly.
Modyn \cite{bother2023towards} introduces a \textit{period-based trigger policy} that performs training periodically and highlights the effectiveness of daily retraining in large-scale environments. In particular, the authors note that the Criteo-TB dataset is organized on a daily basis, making it well-suited for applying time-based training schedules, and use it to conduct daily online training experiments.
we adopted two approaches above in our study as explained in the following Section \ref{section_proposed_recflash_online}.

\begin{algorithm}[t]
\footnotesize
\KwIn{Hash table $\mathsf{HT}$, Trained dataset $\mathcal{D}_{\text{train}} = \{ \langle k, f_k \rangle \}$}
\KwOut{Updated hash table with reassigned physical addresses $a_k$}

\BlankLine %
\textbf{Step 1: Hash table with doubly linked list} \\
Each $k \in \mathsf{HT}$ maintains $(f_k, a_k, \mathsf{prev}(k), \mathsf{next}(k))$

\textbf{Step 2: Determine Threshold key for Remapping} \\
Let $\tau$ be the key satisfying: $f_\tau$ is the top-$x$\%-th value in $\{f_k \mid k \in \mathsf{HT}\}$ \\
Let $\tau_{\text{prev}}$ denote the key immediately preceding $\tau$ in access order

\textbf{Step 3: Insert New Keys Based on Access Frequency} \\
\ForEach{$k_{\text{new}} \in \mathcal{D}_{\text{train}}$}{
    $\texttt{ptr} \gets \mathsf{head}(\mathsf{HT}), \quad \texttt{flag} \gets \texttt{False}$\;
    \While{$\texttt{ptr} \neq \tau$}{
        \If{$f_{k_{\text{new}}} > f_{\texttt{ptr}}$}{
            Insert $k_{\text{new}}$ before $\texttt{ptr}$\;
            Move $\tau$ to the tail of $\mathsf{HT}$\;
            $\tau \gets \tau_{\text{prev}}, \quad \tau_{\text{prev}} \gets \mathsf{prev}(\tau_{\text{prev}})$\;
            $\texttt{flag} \gets \texttt{True}$; \textbf{break}\;
        }
        $\texttt{ptr} \gets \mathsf{next}(\texttt{ptr})$\;
    }
    \If{\texttt{not} flag}{
        Append $k_{\text{new}}$ to $\mathsf{tail}(\mathsf{HT})$
    }
}

\textbf{Step 4: Reassign Physical Addresses $a_k$ } \\
\ForEach{$k \in \mathsf{HT}$}{
    \uIf{$k \in [\mathsf{head}(\mathsf{HT}), \tau]$}{
        $a_k \gets$ reassigned address \tcp*{hot region remapping}
    }
    \uElseIf{$k \in \mathcal{D}_{\text{train}} \land k \in \mathsf{tail}(\mathsf{HT})$}{
        $a_k \gets$ assigned directly \tcp*{not remapped, outside top-$x$\%}
    }
    \Else{
        $a_k \gets a_k$ \tcp*{no change}
    }
}
\caption{Access Frequency-based Adaptive Update in Hash Table}
\label{algo_adapt}
\end{algorithm}

\subsubsection{Low-Overhead Adaptive Remapping in the Online Training}
\label{section_proposed_recflash_online}

Considering the online training, our access frequency-based remapping method leads to two key challenges.
1) Performing frequency-based reordering often requires scanning and repositioning all the entries in the entire hash table, which causes high latency overhead.
2) During the updates, NAND flash requires reading all the data in the block containing the updated data, writing it to new locations, and performing garbage collection by erasing the original block \cite{wu2022joint,li2023ecssd}.
Due to these two challenges, the proposed method significantly impact real-time inference performance with online training.

To help readers intuitively understand how Algorithm \ref{algo_adapt} works, we first briefly explain our low-overhead remapping method using a simple example as follows.
Fig. \ref{Online_Training_Begin} visualizes how access frequencies of new embedding vectors are collected during online inference and the frequencies are compared with a reference hash table to determine when to trigger online training.
During the inference, access frequencies of the embedding vectors for the online training set are recorded in a separate hash table (Fig. \ref{Online_Training_Begin}a) where their access counts are tracked over time.
Afterwards, we can consider two online training policies (\textit{threshold-based trigger policy} and \textit{period-based trigger policy}) as explained in the previous Section \ref{section_proposed_recflash_condition}.
At the end of each day (or any target period), if we use the \textit{period-based trigger policy}, we directly perform the remapping by skipping the following procedure.
Otherwise (\textit{threshold-based trigger policy}), access statistics in the online training set are compared with the reference hash table used for the inference (Fig. \ref{Online_Training_Begin}b) which maintains vector entries sorted in descending order of access frequency.
If the number of vectors in the online training hash table whose access frequency exceeds the top-$x\%$ threshold (\textbf{Hot-item region} in the inference hash table) is greater than pre-defined portion (0.1\% in this example) of the total number of entries in the online training hash table (Fig. \ref{Online_Training_Begin}a), online training is triggered.
Otherwise, we check whether to trigger the online training at the next period (eg., tomorrow).
After the training phase, we can move on the remapping stage.

The \textbf{key idea} behind our low-overhead remapping is to perform the remapping of the embedding vectors at the \textbf{hot-item region} only, instead of reordering all the embedding vectors.
New emerging \textbf{hot items} from the training hast table are selectively inserted into appropriate positions in the reference hash table based on their updated access frequency, and physical addresses are reassigned accordingly.
In this stage, cold vectors from the training hast table and retired \textbf{hot items} from the inference hast table are stored in any free space at the cold-vector region of the inference table.

\begin{figure}[!t]
\centering
\includegraphics[width=0.48\textwidth]{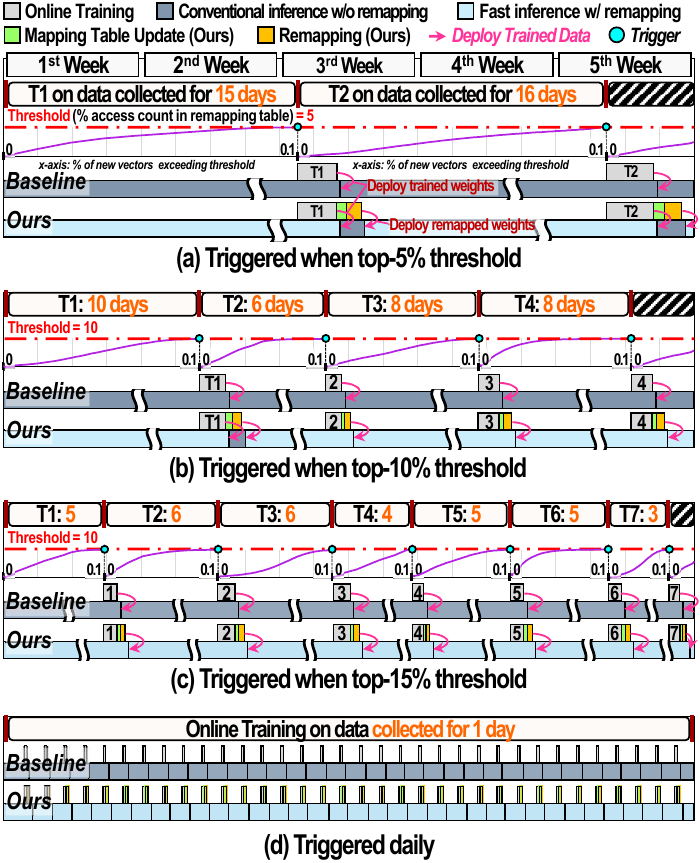}
\vspace{-2mm}
\caption{Overhead of the proposed method under four online training trigger policies: (a) top 5\%, (b) top 10\%, (c) top 15\% access frequency thresholds, and (d) daily trigger using data from the previous day.}
\vspace{-4mm}
\label{Online_Training}
\end{figure}
The insertion logic and address reassignment process are described in detail in Algorithm \ref{algo_adapt}.
Algorithm \ref{algo_adapt} explains how new \textbf{hot items} are efficiently inserted into the existing hash table after online training, and how physical addresses are reassigned for high-access-frequency keys prior to remapping, which is based on the following two points as briefly explained above.
First, maintaining previous and next pointers in each hash table entry allows the access frequency order to be represented as a doubly linked list, enabling efficient reordering by simply updating links without full-table reorganization.
Second, restricting the search range to the top-$x\%$ based on access frequency rather than traversing the entire hash table reduces search overhead and improves update performance.
The algorithm consists of four main steps.
We construct a hash table with a doubly linked list, where each key (vector ID) stores its access count and physical address, along with pointers to its previous and next keys (Step 1).
We identify the \textit{threshold\_key} based on access statistics by selecting the key whose access count ranks in top-$x\%$ (Step 2). The preceding key is also tracked as \textit{threshold\_prev}.
After completing online training on data collected over $y$ days, we sequentially insert all \textit{new keys} from the trained data into the hash table (Step 3).
For each \textit{new key}, we sequentially search from the head of the list up to the \textit{threshold\_key}, comparing the access count of the \textit{new key} with those of existing keys. If the \textit{new key} has a higher access count than the current key, we insert it before the current key, move the existing \textit{threshold\_key} to the tail, and update both the \textit{threshold\_key} and \textit{threshold\_prev} accordingly. If no insertion point is found, the new key is appended at the tail.
After inserting all new keys, we finalize physical address assignments depending on their updated positions in the hash table (Step 4).
For keys from head to \textit{threshold\_key}, positions have changed due to remapping.
Thus, we reassign their physical addresses, which involves reading data from their previous locations, writing to the allocated blocks, and marking the original blocks for erasure.
For the keys inserted at the tail of the list (not in top-$x\%$), there is no need for remapping since these keys are not frequently accessed.
Hence, we simply assign physical addresses to these keys without additional overhead.
Finally, keys that were previously below top-$x\%$ remain unchanged in their positions. Thus, we keep their existing physical addresses, requiring no additional operation.
\begin{figure}[!t]
\vspace{-2mm}
\centering
\includegraphics[width=0.48\textwidth]{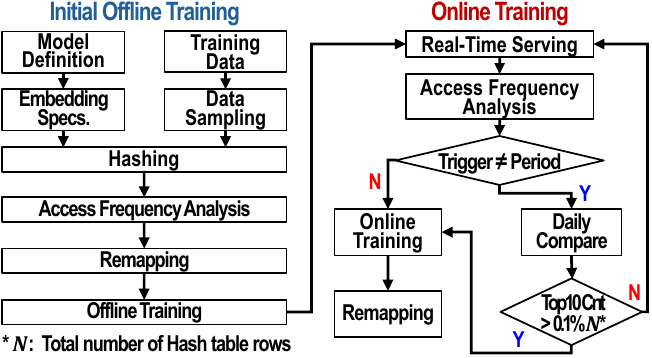}
\caption{Overall Flow of our RecFlash design.}
\vspace{-6mm}
\label{proposed_overflow}
\end{figure}

Meanwhile, as briefly explained before, we need to consider two online training cases: \textit{threshold-based trigger policy} (Fig. \ref{Online_Training}a-c) and \textit{period-based trigger policy} (Fig. \ref{Online_Training}d).
So, we adopt both approaches in our study.
As discussed in the Section \ref{section_intro}, industrial environments often involve real-time inference that continues for several weeks after offline training, with trillions of inferences occurring per day.
Fig. \ref{Online_Training} visualizes the timeline of the proposed method with the two online training schedules during a 5-week simulation period that reflects real-world inference scenario.
Specifically, (Fig. \ref{Online_Training}a) represents a condition where online training is triggered when new accessed vector IDs exceeding the top 5\% access frequency threshold account for more than 0.1\% of the total.
We furthermore evaluated the timeline with 10\% (Fig. \ref{Online_Training}b) and 15\% (Fig. \ref{Online_Training}c) thresholds.
Lastly, (Fig. \ref{Online_Training}d) illustrates a 
period-based trigger policy
scenario, where online training is performed daily using data collected from the previous day.

\begin{figure}[!t]
\vspace{-2mm}
\centering
\includegraphics[width=0.48\textwidth]{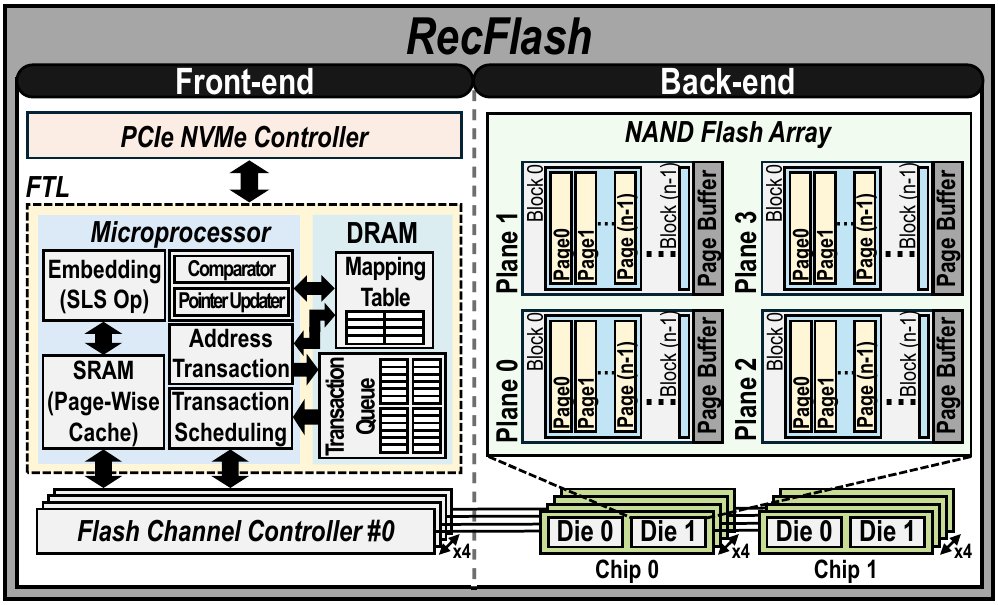}
\vspace{-2mm}
\caption{Top-level architecture of our RecFlash design.}
\vspace{-2mm}
\label{proposed_top}
\end{figure}

\begin{table}[!t]
\renewcommand{\arraystretch}{1.2}
\caption{Benchmark Parameters of DLRM Models.}
\centering
\footnotesize
\begin{tabular}{c|c|c|c|c|c}
\hline
\multirow{2}{*}{\textbf{DLRM}} & \multicolumn{3}{c|}{\textbf{Embedding Layer}} & \multicolumn{2}{c}{\textbf{Fully-Connected Layer}} \\
& Num & Dim & Lookups & Bottom & Top \\ \hline\hline
RMC1  & 8 & 32 & 80 & 128-64-32 & 256-64-1 \\
RMC2  & 32 & 64 & 120 & 256-128-64 & 128-64-1 \\
RMC3  & 10 & 32 & 20 & 2560-1024-256-32 & 512-256-1 \\ \hline
\end{tabular}
\vspace{-2mm}
\label{table_benchmark_DLRM}
\end{table}

\begin{table}[!t]
\renewcommand{\arraystretch}{1.2}
\caption{NAND Flash Configurations Used in Our Evaluation.}
\vspace{-2mm}
\centering
\footnotesize

\begin{tabular*}{\columnwidth}{@{\extracolsep{\fill}} c c c c c c}
\hline
\textbf{Type} & 
\textbf{Page Size} & 
\textbf{\# of Planes} &
\textbf{Latency} &  
\textbf{Energy} & 
\textbf{Die Area} \\
\hline
SLC & 4 KB  & 2 & 25 $\mu$s  & 7.39 $\mu$J   & 89.65 mm$^2$ \\
TLC & 16 KB & 2 & 60 $\mu$s  & 69.06 $\mu$J  & 128.64 mm$^2$ \\
QLC & 16 KB & 2 & 140 $\mu$s & 110.99 $\mu$J & 181.88 mm$^2$ \\
\hline
\end{tabular*}

\vspace{-4mm}
\label{table_nand_config}
\end{table}

To implement our access frequency-based remapping method at the online, we need to carefully analyze and handle the remapping overhead.
The overhead consists of two components: (1) Mapping Table Update, where the hash table is updated based on access frequency using {Algorithm \ref{algo_adapt}}, and (2) Remapping, which involves reading, writing, and erasing data in NAND flash according to the updated mapping table. The first distinction between the baseline and the proposed method appears during the inference phase immediately after offline training.
Fig. \ref{Online_Training} shows the timing diagram of baseline and ours with remapping method.
While the baseline can be deployed immediately after online training, our method incurs the latency with a few hours due to the additional remapping stage.
Before starting the online remapping for our design, we can perform much faster inference with the access frequency-based remapping method. 
However, after finishing online training, it may be concerned that our design can be slower than the baseline due to the remapping latency.
To prevent the performance degradation, right after the online training, we deploy the inference service with the embedding table without proposed remapping method.
In this stage, our performance is identical to the baseline.
After finishing the remapping, we can deploy the faster inference with the remapped hash table than the baseline.
Considering that the remapping time is much smaller than other inference service time, our design shows high throughput compared to the baseline on average by compensating for the overhead.
At the end of each training window, the baseline system must already write the trained embedding tables to the SSD as part of the normal deployment pipeline. 
RecFlash follows the same deployment procedure. 
It then updates the mapping table and performs remapping before the remapped tables become active, so the extra DRAM activity and controller work are confined to this infrequent deployment phase. 
Consequently, RecFlash incurs only a small and bounded overhead during the infrequent deployment and remapping phase, including mapping table updates and the associated NAND writes. 
Steady state inference then proceeds without additional remapping induced latency. 
Since these extra writes occur only at deployment time and are limited in scope, they are unlikely to materially affect SSD lifetime.

\subsubsection{Overall Flow of Proposed Algorithm}
\label{section_proposed_recflash_algo}

The overall flow of the proposed algorithm (Fig. \ref{proposed_overflow}) starts with the offline initial training stage.
We first sweep the sampled input training data, and the access counts for each item are evaluated.
Based on the frequency evaluation, we construct the hash table with access frequency-based descending ordered embedding items.
Then, we start to train the embedding vectors.
In this time, the physical addresses of the embedding vectors correspond to the hash table we developed.
After the training is complete, we start to service the inference task.

During the inference phase, we need to perform the online training to reflect the users' behavior at the recommendation system.
As explained earlier, we can select one of two well-known online training approaches: \textit{threshold-based trigger policy} and \textit{period-based trigger policy}.
If we use the \textit{period-based trigger policy}, we perform the online training and remapping of the embedding vectors daily.
Otherwise (\textit{period-based trigger policy}), we check the training trigger condition every day as explained in the earlier Section.
In this case, during the \textbf{hot item} remapping stage, we service the inference with the baseline hash table without remapping to prevent it from losing the accuracy.
After finishing the remapping, we can use the remapped hash table to increase the throughput.

\subsubsection{Hardware Design}
\label{section_proposed_recflash_hardware}

Fig. \ref{proposed_top} illustrates the top-level architecture of RecFlash, which consists of a front-end SSD controller and a NAND flash-based back-end. The front-end includes a PCIe NVMe controller for host communication, a microprocessor that handles FTL logic and embedding operations, and DRAM that stores the mapping table. The back-end follows a conventional NAND flash structure, consisting of multiple channels connected to chips composed of dies, planes, blocks, and pages.
To support adaptive remapping, we designed and synthesized a lightweight hardware module that implements the mapping table update logic described in Algorithm \ref{algo_adapt}. 
This module resides in the FTL and directly interacts with the mapping table stored in DRAM.
It maintains the access frequency order by utilizing the prev/next pointers between hash table entries and operates by traversing only in the \textbf{hot-item region} and inserting the target key at the appropriate position. 
This approach enables low-latency updates without requiring full-table reordering, thereby minimizing the overhead of remapping.
The logic consists of two key hardware blocks: a comparator and a pointer updater. 
The comparator determines whether the access count of a given key exceeds the threshold to enter the \textbf{hot-item region}, and the pointer updater modifies the prev/next pointers of the surrounding entries to maintain the order in the hash table. 
The search range is limited during insertion to reduce computational overhead, and we measured the execution time of this update logic under realistic access patterns through RTL-level simulation.
To quantify the silicon overhead of the proposed logic, we synthesized the comparator and pointer-updater blocks using a 28\,nm CMOS standard-cell library at 500\,MHz. The comparator occupies $5.63\times 10^{-5}\,\text{mm}^2$ and the pointer updater $9.20\times 10^{-5}\,\text{mm}^2$, for a total of $14.82\times 10^{-5}\,\text{mm}^2$. Also, their per-operation energy is $6.39\times 10^{-9}\,\mu\text{J}$ and $9.94\times 10^{-9}\,\mu\text{J}$, respectively.
Even with process-node differences, the results show that the hardware cost of the comparator and pointer-updater blocks is negligible in both area and energy compared to the back-end NAND flash configurations in Table \ref{table_nand_config}.
RecFlash was evaluated by modifying the latest SSD simulator, MQSim\cite{tavakkol2018mqsim}. 
Similar to RecSSD and RM-SSD, embedding operations are performed in the SSD’s FTL where data is retrieved from the NAND flash array and SLS computations are processed in the FTL.
We modified the FTL firmware to implement the data remapping based on access frequency, thereby maximizing the utilization of the page buffer. Furthermore, we added a page-level cache in the FTL to enable fast access to frequently used embedding vectors.

\subsubsection{Related Work}
\label{section_prelim_related}

SOML \cite{liu2019soml} revisits how a read operation is defined in 3D-NAND SSDs. 
Today, even if the host requests only a few cache lines, the flash chip must read an entire page (16 KB) at a time, which wastes both time and energy. 
It addresses this mismatch by introducing a new read operation that can fetch only the needed small portions of data, and even gather them from multiple rows within a plane in a single read.
To enable this fine-grained read behavior, it adds extra selector structures and peripheral support inside the 3D-NAND chip and relies on a controller and scheduler that support SOML read operations.

Our work is also motivated by the mismatch between small logical reads and large page-level reads, but targets a different design point. 
RecFlash assumes commodity 3D NAND devices that only support conventional page-level read operations and does not modify the NAND cell array or peripheral circuits.
Instead, RecFlash focuses on recommendation workloads and proposes an embedding-aware in-storage accelerator that optimizes the data layout through access-frequency-based remapping, plane-level distribution, and page-wise caching on top of the existing FTL and page-buffer organization.
These techniques reduce the number of page read operations ($t_{R}$) and improve page-buffer and intra-plane parallelism for embedding-heavy recommendation models, without requiring any changes to the NAND device architecture.
In summary, both SOML and RecFlash start from the same observation about the mismatch between small logical reads and large page-level reads, but RecFlash tackles this problem by reorganizing the data layout and system-level mechanisms, rather than by modifying the underlying hardware.

\section{Results}
\label{section_result}

\subsection{Experimental Setup}
\label{section_result_setup}

\begin{figure*}[t]
\centering
\includegraphics[width=0.98\textwidth]{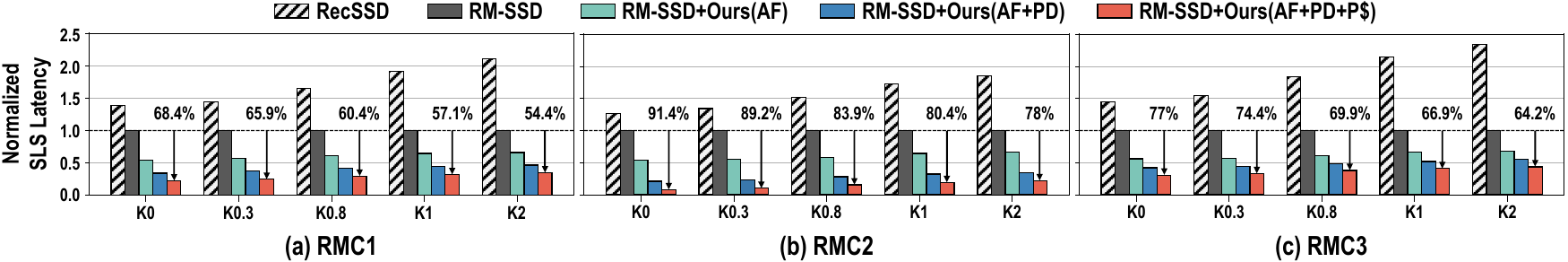}
\vspace{-2mm}
\caption{Normalized embedding operation latency for TLC memory configuration when sweeping trace-based datasets (K0-K2) across three different DLRM models: (a) RMC1, (b) RMC2, and (c) RMC3.}
\vspace{-2mm}
\label{result1}
\end{figure*}
\begin{figure*}[t]
\centering
\includegraphics[width=0.98\textwidth]{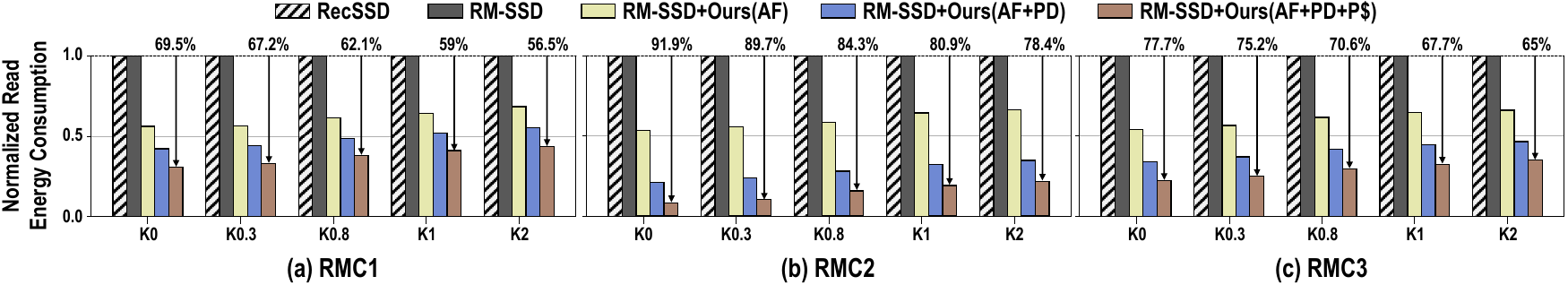}
\vspace{-2mm}
\caption{Normalized read energy consumption for TLC memory configuration when sweeping trace-based datasets (K0-K2) across three different DLRM models: (a) RMC1, (b) RMC2, and (c) RMC3.}
\vspace{-2mm}
\label{result2}
\end{figure*}

We used the Facebook’s DLRM models as a benchmark which include both embedding-dominated and fully connected layer-dominated models, as detailed in Table \ref{table_benchmark_DLRM}.
Similar to the baseline, we used a synthetic trace generator to control the locality level ($K$). The parameter K was set to 0, 0.3, 0.8, 1, and 2, corresponding to unique access rates ranging from 8\% to 66\%.
Like RecSSD, we assumed each embedding table contains 1 million rows and that the DRAM cache stores up to 2K vectors per table. The SSD’s internal DRAM was used only for mapping tables, and as in RM-SSD, we excluded DRAM caching due to its limited benefit in RecSSD.
To evaluate RecFlash, we used datasets with varying locality and compared execution time and read energy consumption across SLC, TLC, and QLC types against RecSSD and RM-SSD.
For the three NAND flash configurations (8Gb SLC \cite{dong2012nvsim}, 512Gb TLC \cite{kim2017512}, and 1Tb QLC \cite{lee20181tb}), we adopt the device parameters (page size, number of planes per die, page read latency ($t_R$), page read energy, and die area) summarized in Table \ref{table_nand_config}.
Energy was measured using NVSim \cite{dong2012nvsim} and 3D-FPIM \cite{lee20223d} with these memory configurations.
We also tested RecFlash on the Criteo Terabyte dataset to reflect real-world deployment. In the first experiment, we trained the model on day0–22 and evaluated inference performance on day23 (static). In the second, we simulated real-time inference with simultaneous online training after an initial short training phase. The day-wise structure of the dataset enabled assessment of online training trigger policies described in Section \ref{section_proposed_recflash_online}.

\subsection{Results}
\label{section_result_result}

\begin{figure*}[t]
\centering
\includegraphics[width=0.98\textwidth]{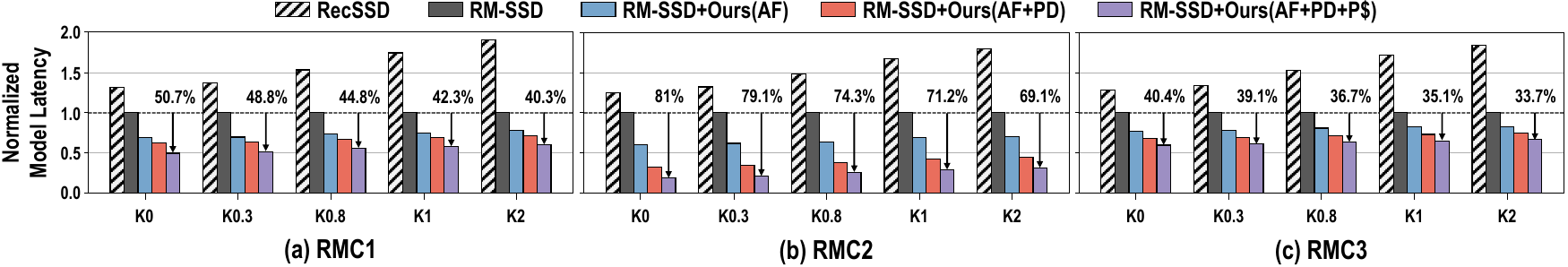}
\vspace{-2mm}
\caption{Normalized end-to-end model latency for TLC memory configuration when sweeping trace-based datasets (K0-K2) across three different DLRM models: (a) RMC1, (b) RMC2, and (c) RMC3.}
\vspace{-4mm}
\label{result3}
\end{figure*}
\begin{figure}[t]
\centering
\includegraphics[width=0.48\textwidth]{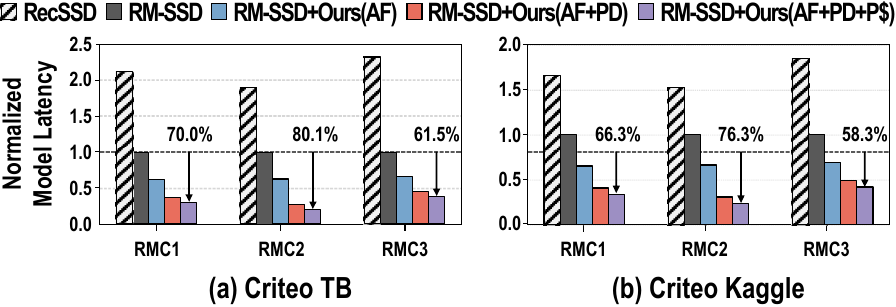}
\vspace{-2mm}
\caption{Normalized end-to-end model latency for TLC memory configuration when evaluating three different DLRM models (RMC1–RMC3) on two real-world datasets: (a) Criteo TB and (b) Criteo Kaggle.}
\vspace{-6mm}
\label{result4}
\end{figure}

Fig. \ref{result1} compares the normalized embedding operation latency of RecSSD, RM-SSD, and the proposed RecFlash design across three DLRM models (RMC1, RMC2, and RMC3) with varying locality ($K0$-$K2$) on TLC-based SSDs. As dataset locality increases (lower $K$ value), RecFlash shows significant latency improvements compared to RM-SSD.
This is due to the proposed remapping method, which enhances data reuse in the page buffer and maximizes page-wise cache utilization in high-locality datasets. In particular, the RMC2 model, where embedding operations are critical, demonstrates a latency reduction of 78\% to 91.4\% with RecFlash compared to RM-SSD, marking the greatest improvement. The RMC1 and RMC3 models also show notable performance gains, with improvements ranging from 54.4\% to 68.4\% and 64.2\% to 77\%, respectively. 
We also evaluated performance across different memory configurations beyond TLC. On average, RecFlash achieved latency reductions of approximately 54\%, 77\%, and 62\% for RMC1, RMC2, and RMC3, respectively, in SLC-based SSDs. In QLC-based SSDs, the reductions were around 66\%, 89\%, and 75\%, respectively. These results demonstrate that the RecFlash design consistently enhances performance across various memory configurations, DLRM models, and datasets with different locality characteristics.

Fig. \ref{result2} compares the normalized memory read energy consumption for RecSSD, RM-SSD, and RecFlash in TLC memory configurations. NVSim and 3D-FPIM simulator were modified to measure energy consumption during memory read operations for each model. As discussed in Section \ref{section_proposed_prev}, both RecSSD and RM-SSD take the same amount of time to load data from the memory array to the page buffer. However, the key difference came from how data is accessed—sequentially or randomly—from the page buffer. As a result, the energy consumption during memory reads is the same for RecSSD and RM-SSD. RecFlash, however, significantly reduces energy consumption, especially in high-locality datasets. The RMC2 model showed the largest reduction, with energy savings of up to 91.9\%, while the RMC1 and RMC3 models showed reductions of up to 69.5\% and 77.7\%, respectively. 

Fig. \ref{result3} presents a comparison of the normalized end-to-end model latency for RecSSD, RM-SSD, and RecFlash in TLC-based SSDs. As dataset locality increases, RecFlash consistently outperforms both RecSSD and RM-SSD, with performance improvements of up to 50.7\% in RMC1, 81\% in RMC2, and 40.4\% in RMC3. 
The proposed method achieved the most significant latency reduction in the RMC2 model, where embedding operations dominate. In contrast, the performance gain in the RMC3 model was relatively limited due to the high proportion of MLP computations.

\begin{figure*}[t]
\vspace{-1mm}
\centering
\includegraphics[width=0.98\textwidth]{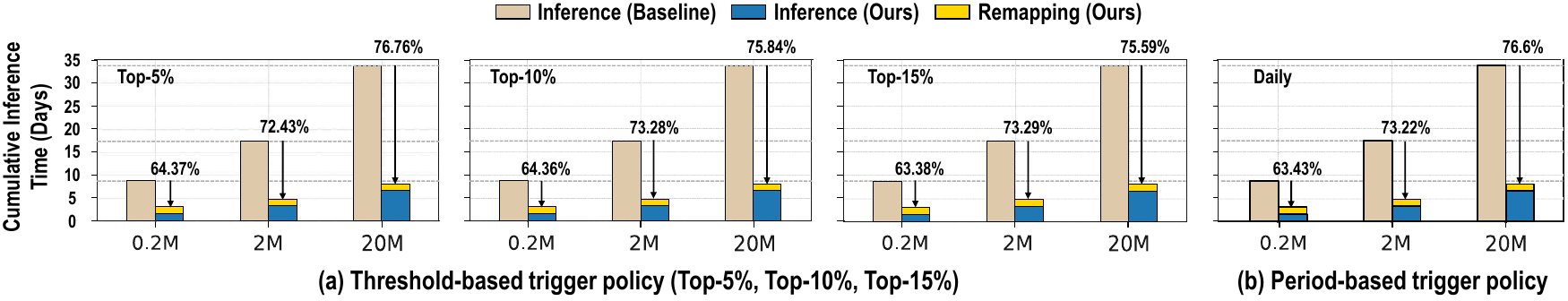}
\vspace{-2mm}
\caption{Cumulative inference time (in days) under different online training trigger policies by sweeping daily inference count from 0.2M to 20M: (a) threshold-based trigger policy (Top-5\%, Top-10\%, Top-15\%) and (b) period-based trigger policy (daily).}
\vspace{-4mm}
\label{result5}
\end{figure*}

Fig. \ref{result4} presents a comparison of normalized end-to-end model latency among RecSSD, RM-SSD, and RecFlash when evaluating three DLRM models (RMC1–RMC3) on two real-world datasets: Criteo TB and Criteo Kaggle. In the Criteo TB experiments, which serve as the primary evaluation in our study, RecFlash reduces latency by up to 70.0\% for RMC1, 80.1\% for RMC2, and 61.5\% for RMC3 compared to RM-SSD. These results demonstrate that RecFlash is highly effective in handling large-scale datasets where embedding vector reuse is significant. To validate the generality of our approach, we further conducted experiments on the widely used Criteo Kaggle dataset, which consists of six days of click log data. As shown in (Fig. \ref{Online_Training}b), RecFlash again exhibits consistent improvements, reducing latency by 66.3\% for RMC1, 76.3\% for RMC2, and 58.3\% for RMC3 over RM-SSD. Although the Kaggle dataset is smaller than Criteo TB, the results confirm that RecFlash’s frequency-aware remapping and page-wise caching are effective even in mid-scale settings.

So far, we evaluated the performance of embedding-optimized data remapping with weights trained offline.
We furthermore analyze the effectiveness of data remapping with online training case.
Fig. \ref{result5} shows the cumulative inference time over 35 days under different online training trigger policies, assuming a single inference machine and sweeping the number of daily inferences from 0.2M to 20M. The baseline is RM-SSD, while the proposed RecFlash adopts a dedicated embedding accelerator and access-frequency-based remapping. As a result, RecFlash already achieves performance benefits from the very first inference stage, even before online training begins.
The baseline represents the total inference time accumulated over 35 days with online training, whereas RecFlash accumulates the inference time measured including each online training and remapping phase. Since online training is performed concurrently with inference, training time is excluded from the cumulative inference time. In contrast, the remapping process introduces temporary service delay and is thus explicitly included as an overhead, shown in yellow in the figure.
Although retraining at shorter periods is generally expected to increase remapping overhead and degrade performance, RecFlash minimizes such overhead by limiting the remapping to only the hot-item region of the hash table through its adaptive remapping mechanism. 
Furthermore, when retraining occurs more frequently, less data is collected each day.
This results in a smaller online training hash table and fewer vectors that need to be inserted or remapped, which ultimately lowers the remapping cost.
Through this strategy, RecFlash achieves significant performance improvements across all trigger policies and inference counts. Notably, with 20 million inferences per day, RecFlash achieves up to 76.7\% reduction in cumulative inference time compared to the baseline.

\section{Conclusion}
\label{section_conclusion}
This paper presented RecFlash, a fast and energy-efficient recommendation inference accelerator based on NAND flash in-storage computing (ISC). To address the internal bandwidth underutilization caused by the irregular memory access patterns in embedding layers, we optimized the data layout by combining access frequency-based remapping, plane-level distribution, and page-wise caching techniques. These methods improve page buffer utilization and maximize the parallelism of the flash memory array, ultimately enhancing inference performance.
To further improve long-term inference efficiency, we introduced an online training-aware remapping strategy that supports both threshold-based and period-based trigger policies. In particular, to minimize the overhead typically associated with remapping, we proposed an adaptive remapping technique that selectively updates only the hot-item region of the hash table. This approach enables efficient updates without requiring global table reordering, thereby minimizing disruptions to real-time inference.
Experimental results show that RecFlash reduces inference latency by up to 81\%, energy consumption by 91.9\%, and cumulative inference time with online training by up to 76.7\%, compared to RM-SSD and RecSSD baselines.


\appendices
\section{Major Extensions over the Conference Version}
\label{appendices}

Compared to the conference version published in APCCAS 2025, several major updates and extensions have been added \textcolor{black}{to the extended submission}:

\begin{enumerate}
    \item Compared to the conference version, this \textcolor{black}{extended} paper additionally addresses online training by introducing an adaptive remapping algorithm that operates during runtime. Specifically, Algorithm~1 selectively updates only the hot region of the hash table by inserting frequently accessed keys into the top positions and reassigning their physical addresses, thereby minimizing the remapping overhead (Section \ref{section_proposed_recflash_condition}, \ref{section_proposed_recflash_online} / Fig. \ref{Online_Training_Begin} / Algorithm \ref{algo_adapt}).
   
    \item We evaluate two types of online training triggers: access frequency threshold-based policies and daily triggers. To mitigate the latency overhead from remapping after online training, we optionally support a two-stage deployment strategy—initially deploying the trained weights, then switching to remapped weights after remapping completes—only when necessary to avoid service delays (Section \ref{section_proposed_recflash_online} / Fig. \ref{Online_Training}).

    \item To support adaptive remapping, we implemented a lightweight hardware module inside the FTL that updates the DRAM-based mapping table based on Algorithm~1. It consists of a comparator to check if a key exceeds the hot-item threshold and a pointer updater to adjust the \texttt{prev}/\texttt{next} pointers in the hash table. This localized update mechanism enables low-latency remapping without full-table reordering (Section \ref{section_proposed_recflash_hardware} / Fig. \ref{proposed_top}).
    
    \item We follow a two-phase flow consisting of offline and online training. In the offline phase, we sort embedding vectors by access frequency, train them accordingly, and begin inference using the resulting hash table. During the online phase, we selectively trigger training and remapping based on threshold- or period-based policies, updating the system only when necessary (Section \ref{section_proposed_recflash_algo} / Fig. \ref{proposed_overflow}).
    
    \item Compared to the conference version, this \textcolor{black}{extended} paper extends the evaluation scope by separately analyzing embedding latency, memory read energy, and end-to-end model latency using NVSim and 3D-FPIM. It also evaluates the design on two real-world datasets (Criteo TB and Kaggle), whereas the conference version only reported total inference results on synthetic workloads (Section \ref{section_result_setup}, \ref{section_result_result} / Fig. \ref{result1}, \ref{result2}, \ref{result4}).
 
    \item Unlike the conference version that only considered the offline phase, we additionally evaluate online training scenarios using both threshold-based and period-based triggers. By remapping only hot items and overlapping training with inference while counting only remapping as overhead, we reduce cumulative inference time effectively (Section \ref{section_result_setup}, \ref{section_result_result} / Fig. \ref{result5}).
\end{enumerate}

\bibliographystyle{IEEEtran}
\bibliography{TC_V1/recflash_references}

@inproceedings{he2017neural,
  title={Neural collaborative filtering},
  author={He, Xiangnan and Liao, Lizi and Zhang, Hanwang and Nie, Liqiang and Hu, Xia and Chua, Tat-Seng},
  booktitle={Proceedings of the 26th international conference on world wide web},
  pages={173--182},
  year={2017}
}

@inproceedings{mooney2000content,
  title={Content-based book recommending using learning for text categorization},
  author={Mooney, Raymond J and Roy, Loriene},
  booktitle={Proceedings of the fifth ACM conference on Digital libraries},
  pages={195--204},
  year={2000}
}

@article{naumov2019deep,
  title={Deep learning recommendation model for personalization and recommendation systems},
  author={Naumov, Maxim and Mudigere, Dheevatsa and Shi, Hao-Jun Michael and Huang, Jianyu and Sundaraman, Narayanan and Park, Jongsoo and Wang, Xiaodong and Gupta, Udit and Wu, Carole-Jean and Azzolini, Alisson G and others},
  journal={arXiv preprint arXiv:1906.00091},
  year={2019}
}

@inproceedings{hazelwood2018applied,
  title={Applied machine learning at facebook: A datacenter infrastructure perspective},
  author={Hazelwood, Kim and Bird, Sarah and Brooks, David and Chintala, Soumith and Diril, Utku and Dzhulgakov, Dmytro and Fawzy, Mohamed and Jia, Bill and Jia, Yangqing and Kalro, Aditya and others},
  booktitle={2018 IEEE International Symposium on High Performance Computer Architecture (HPCA)},
  pages={620--629},
  year={2018},
  organization={IEEE}
}

@inproceedings{kim2023recpim,
  title={RecPIM: A PIM-Enabled DRAM-RRAM Hybrid Memory System For Recommendation Models},
  author={Kim, Heewoo and Ye, Haojie and Mudge, Trevor and Dreslinski, Ronald and Talati, Nishil},
  booktitle={2023 IEEE/ACM International Symposium on Low Power Electronics and Design (ISLPED)},
  pages={1--6},
  year={2023},
  organization={IEEE}
}

@inproceedings{kwon2019tensordimm,
  title={Tensordimm: A practical near-memory processing architecture for embeddings and tensor operations in deep learning},
  author={Kwon, Youngeun and Lee, Yunjae and Rhu, Minsoo},
  booktitle={Proceedings of the 52nd Annual IEEE/ACM International Symposium on Microarchitecture},
  pages={740--753},
  year={2019}
}

@inproceedings{ke2020recnmp,
  title={Recnmp: Accelerating personalized recommendation with near-memory processing},
  author={Ke, Liu and Gupta, Udit and Cho, Benjamin Youngjae and Brooks, David and Chandra, Vikas and Diril, Utku and Firoozshahian, Amin and Hazelwood, Kim and Jia, Bill and Lee, Hsien-Hsin S and others},
  booktitle={2020 ACM/IEEE 47th Annual International Symposium on Computer Architecture (ISCA)},
  pages={790--803},
  year={2020},
  organization={IEEE}
}

@inproceedings{kal2021space,
  title={Space: locality-aware processing in heterogeneous memory for personalized recommendations},
  author={Kal, Hongju and Lee, Seokmin and Ko, Gun and Ro, Won Woo},
  booktitle={2021 ACM/IEEE 48th Annual International Symposium on Computer Architecture (ISCA)},
  pages={679--691},
  year={2021},
  organization={IEEE}
}

@inproceedings{liu2023accelerating,
  title={Accelerating Personalized Recommendation with Cross-level Near-Memory Processing},
  author={Liu, Haifeng and Zheng, Long and Huang, Yu and Liu, Chaoqiang and Ye, Xiangyu and Yuan, Jingrui and Liao, Xiaofei and Jin, Hai and Xue, Jingling},
  booktitle={Proceedings of the 50th Annual International Symposium on Computer Architecture},
  pages={1--13},
  year={2023}
}

@electronic{criteo-terabyte,
 author   = {Criteo},
 title     = {Criteo AI Labs Ad Terabyte},
 url       = {https://labs.criteo.com/2013/12/downloadterabyte-click-logs/}
}

@inproceedings{wilkening2021recssd,
  title={RecSSD: near data processing for solid state drive based recommendation inference},
  author={Wilkening, Mark and Gupta, Udit and Hsia, Samuel and Trippel, Caroline and Wu, Carole-Jean and Brooks, David and Wei, Gu-Yeon},
  booktitle={Proceedings of the 26th ACM International Conference on Architectural Support for Programming Languages and Operating Systems},
  pages={717--729},
  year={2021}
}

@inproceedings{sun2022rm,
  title={Rm-ssd: In-storage computing for large-scale recommendation inference},
  author={Sun, Xuan and Wan, Hu and Li, Qiao and Yang, Chia-Lin and Kuo, Tei-Wei and Xue, Chun Jason},
  booktitle={2022 IEEE International Symposium on High-Performance Computer Architecture (HPCA)},
  pages={1056--1070},
  year={2022},
  organization={IEEE}
}

@inproceedings{khakifirooz202130,
  title={30.2 a 1tb 4b/cell 144-tier floating-gate 3d-nand flash memory with 40mb/s program throughput and 13.8 gb/mm 2 bit density},
  author={Khakifirooz, Ali and Balasubrahmanyam, Sriram and Fastow, Richard and Gaewsky, Kristopher H and Ha, Chang Wan and Haque, Rezaul and Jungroth, Owen W and Law, Steven and Madraswala, Aliasgar S and Ngo, Binh and others},
  booktitle={2021 IEEE International Solid-State Circuits Conference (ISSCC)},
  volume={64},
  pages={424--426},
  year={2021},
  organization={IEEE}
}

@inproceedings{park202130,
  title={30.1 A 176-Stacked 512Gb 3b/Cell 3D-NAND Flash with 10.8 Gb/mm 2 Density with a Peripheral Circuit Under Cell Array Architecture},
  author={Park, Jae-Woo and Kim, Doogon and Ok, Sunghwa and Park, Jaebeom and Kwon, Taeheui and Lee, Hyunsoo and Lim, Sungmook and Jung, Sun-Young and Choi, Hyeongjin and Kang, Taikyu and others},
  booktitle={2021 IEEE International Solid-State Circuits Conference (ISSCC)},
  volume={64},
  pages={422--423},
  year={2021},
  organization={IEEE}
}

@inproceedings{kim202328,
  title={28.2 A High-Performance 1Tb 3b/Cell 3D-NAND Flash with a 194MB/s Write Throughput on over 300 Layers},
  author={Kim, Bvunarvul and Lee, Seungpil and Hah, Beomseok and Park, Kanawoo and Park, Yongsoon and Jo, Kangwook and Noh, Yujong and Seol, Hyeoncheon and Lee, Hyunsoo and Shin, Jaehyeon and others},
  booktitle={2023 IEEE International Solid-State Circuits Conference (ISSCC)},
  pages={27--29},
  year={2023},
  organization={IEEE}
}

@inproceedings{huh202013,
  title={13.2 a 1tb 4b/cell 96-stacked-wl 3d nand flash memory with 30mb/s program throughput using peripheral circuit under memory cell array technique},
  author={Huh, Hwang and Cho, Wanik and Lee, Jinhaeng and Noh, Yujong and Park, Yongsoon and Ok, Sunghwa and Kim, Jongwoo and Cho, Kayoung and Lee, Hyunchul and Kim, Geonu and others},
  booktitle={2020 IEEE International Solid-State Circuits Conference-(ISSCC)},
  pages={220--221},
  year={2020},
  organization={IEEE}
}

@Manual{ltc3600,
  title        = {3V, 8G-bit NAND Flash Memory},
  year         = {2022},
  number       = {MX60LF8G28AD},
  note         = {Rev. 1.1.},
  organization = {Macronix},
  url          = {https://www.macronix.com},
}

@inproceedings{wang2024rap,
  title={Rap: Resource-aware automated gpu sharing for multi-gpu recommendation model training and input preprocessing},
  author={Wang, Zheng and Wang, Yuke and Deng, Jiaqi and Zheng, Da and Li, Ang and Ding, Yufei},
  booktitle={Proceedings of the 29th ACM International Conference on Architectural Support for Programming Languages and Operating Systems, Volume 2},
  pages={964--979},
  year={2024}
}

@inproceedings{lai2023adaembed,
  title={$\{$AdaEmbed$\}$: Adaptive embedding for $\{$Large-Scale$\}$ recommendation models},
  author={Lai, Fan and Zhang, Wei and Liu, Rui and Tsai, William and Wei, Xiaohan and Hu, Yuxi and Devkota, Sabin and Huang, Jianyu and Park, Jongsoo and Liu, Xing and others},
  booktitle={17th USENIX Symposium on Operating Systems Design and Implementation (OSDI 23)},
  pages={817--831},
  year={2023}
}

@inproceedings{bother2023towards,
  title={Towards a platform and benchmark suite for model training on dynamic datasets},
  author={B{\"o}ther, Maximilian and Strati, Foteini and Gsteiger, Viktor and Klimovic, Ana},
  booktitle={Proceedings of the 3rd Workshop on Machine Learning and Systems},
  pages={8--17},
  year={2023}
}

@inproceedings{wu2022joint,
  title={A joint management middleware to improve training performance of deep recommendation systems with SSDs},
  author={Wu, Chun-Feng and Wu, Carole-Jean and Wei, Gu-Yeon and Brooks, David},
  booktitle={Proceedings of the 59th ACM/IEEE Design Automation Conference},
  pages={157--162},
  year={2022}
}

@inproceedings{li2023ecssd,
  title={Ecssd: Hardware/data layout co-designed in-storage-computing architecture for extreme classification},
  author={Li, Siqi and Tu, Fengbin and Liu, Liu and Lin, Jilan and Wang, Zheng and Kang, Yangwook and Ding, Yufei and Xie, Yuan},
  booktitle={Proceedings of the 50th annual international symposium on computer architecture},
  pages={1--14},
  year={2023}
}

@inproceedings{tavakkol2018mqsim,
  title={$\{$MQSim$\}$: A framework for enabling realistic studies of modern $\{$Multi-Queue$\}$$\{$SSD$\}$ devices},
  author={Tavakkol, Arash and G{\'o}mez-Luna, Juan and Sadrosadati, Mohammad and Ghose, Saugata and Mutlu, Onur},
  booktitle={16th USENIX Conference on File and Storage Technologies (FAST 18)},
  pages={49--66},
  year={2018}
}

@article{dong2012nvsim,
  title={Nvsim: A circuit-level performance, energy, and area model for emerging nonvolatile memory},
  author={Dong, Xiangyu and Xu, Cong and Xie, Yuan and Jouppi, Norman P},
  journal={IEEE Transactions on Computer-Aided Design of Integrated Circuits and Systems},
  volume={31},
  number={7},
  pages={994--1007},
  year={2012},
  publisher={IEEE}
}

@inproceedings{lee20223d,
  title={3D-FPIM: An Extreme Energy-Efficient DNN Acceleration System Using 3D NAND Flash-Based In-Situ PIM Unit},
  author={Lee, Hunjun and Kim, Minseop and Min, Dongmoon and Kim, Joonsung and Back, Jongwon and Yoo, Honam and Lee, Jong-Ho and Kim, Jangwoo},
  booktitle={2022 55th IEEE/ACM International Symposium on Microarchitecture (MICRO)},
  pages={1359--1376},
  year={2022},
  organization={IEEE}
}

@article{kim2017512,
  title={A 512-Gb 3-b/cell 64-stacked WL 3-D-NAND flash memory},
  author={Kim, Chulbum and Kim, Doo-Hyun and Jeong, Woopyo and Kim, Hyun-Jin and Park, Il Han and Park, Hyun-Wook and Lee, JongHoon and Park, JiYoon and Ahn, Yang-Lo and Lee, Ji Young and others},
  journal={IEEE Journal of Solid-State Circuits},
  volume={53},
  number={1},
  pages={124--133},
  year={2017},
  publisher={IEEE}
}

@inproceedings{lee20181tb,
  title={A 1Tb 4b/cell 64-stacked-WL 3D NAND flash memory with 12MB/s program throughput},
  author={Lee, Seungjae and Kim, Chulbum and Kim, Minsu and Joe, Sung-min and Jang, Joonsuc and Kim, Seungbum and Lee, Kangbin and Kim, Jisu and Park, Jiyoon and Lee, Han-Jun and others},
  booktitle={2018 IEEE International Solid-State Circuits Conference-(ISSCC)},
  pages={340--342},
  year={2018},
  organization={IEEE}
}

@inproceedings{baik2025recflash,
  title        = {RecFlash: Fast Recommendation Inference on NAND Flash-Based In-Storage Computing with Embedding-Optimized Data Mapping},
  author       = {Baik, Jangho and Ji, Gisan and Shim, Wonbo and Ryu, Sungju},
  booktitle    = {Proceedings of the IEEE Asia Pacific Conference on Circuits and Systems (APCCAS)},
  year         = {2025},
  note         = {to be published}
}

@inproceedings{liu2019soml,
  title={SOML read: Rethinking the read operation granularity of 3D NAND SSDs},
  author={Liu, Chun-Yi and Kotra, Jagadish B and Jung, Myoungsoo and Kandemir, Mahmut T and Das, Chita R},
  booktitle={Proceedings of the Twenty-Fourth International Conference on Architectural Support for Programming Languages and Operating Systems},
  pages={955--969},
  year={2019}
}

\newpage

\vfill

\end{document}